\documentclass[useAMS,usenatbib,fleqn]{mn2e}
\usepackage{amsmath,amssymb}
\usepackage{bm}
\usepackage[dvipdfmx]{graphicx}
\usepackage{wrapfig}
\usepackage{here}
\usepackage{float}
\usepackage{fancybox}
\usepackage{color}

\title[Mergers of accreting stellar-mass black holes]
{Mergers of accreting stellar-mass black holes}
\author[H. Tagawa, M. Umemura \& N. Gouda]{H. Tagawa$^{1,2}$\thanks{E-mail:
email@address ; tagawahr@nao.ac.jp }, M.Umemura$^{3}$, and N. Gouda$^{1,2}$\\
$^{1}$The University of Tokyo, 7-3-1 Hongo Bunkyo, Tokyo 113-0033, Japan\\
$^{2}$National Astronomical Observatory of Japan, 2-21-1 Osawa, Mitaka, Tokyo
181-8588, Japan\\
$^{3}$ Center for Computational Sciences, University of Tsukuba, Tsukuba, Ibaraki 305-8577, Japan}
\begin{document}

\date{28 July 2016}

\pagerange{\pageref{firstpage}--\pageref{lastpage}} \pubyear{2016}

\maketitle

\label{firstpage}

\begin{abstract}

We present post-Newtonian $N$-body simulations on mergers of accreting stellar-mass black holes (BHs),
where such general relativistic effects as the pericenter shift and gravitational wave (GW) emission 
are taken into consideration.
The attention is concentrated on the effects of the dynamical friction and the Hoyle-Lyttleton mass 
accretion by ambient gas.
We consider a system composed of ten BHs with initial mass of $30~M_\odot$. 
As a result, we show that mergers of accreting stellar-mass BHs are
classified into four types: a gas drag-driven, an interplay-driven, a three body-driven, 
or an accretion-driven merger.
We find that BH mergers proceed before significant mass accretion, 
even if the accretion rate is $\sim10$ Eddington accretion rate, 
and then all BHs can merge into one heavy BH. 
Using the simulation results for a wide range of parameters,
we derive a critical accretion rate ($\dot{m}_{\rm c}$), below which the BH growth is 
promoted faster by mergers.
Also, it is found that the effect of the recoil by the GW emission can reduce $\dot{m}_{\rm c}$ especially
in gas number density higher than $10^8~{\rm cm}^{-3}$, and enhance the escape
probability of merged BHs. 
Very recently, a gravitational wave event, GW150914, as a result of 
the merger of a $\sim 30~M_\odot$ BH binary 
has been detected 
\citep{abbott16}. 
Based on the present simulations, the BH merger in GW150914 is
likely to be driven by three-body encounters accompanied by a few $M_\odot$ of gas accretion,
in high-density environments like dense interstellar clouds or galactic nuclei.

\end{abstract}


\section{INTRODUCTION}

Recent observations have revealed the existence of supermassive black holes (SMBHs) with masses $\gtrsim10^9~M_\odot$ 
at redshifts higher than 6 \citep*{fan01,kur07,mor11,wu15}.
However, the formation history of these SMBHs is not still revealed. 
There are two major competitive scenarios for the growth of SMBHs: one is the mass accretion, and the other is the merger of BHs
(or stars) \citep*{vol12,hai13}. 
As for the mass accretion, the constraints from observed SMBHs at high redshifts have been argued. 
Possible building blocks of SMBHs are the remnants of first stars. 
First stars of several tens $M_\odot$ can leave black holes (BHs) of few tens $M_\odot$ 
after supernova explosion \citep{heg02}. 
If recently discovered high-redshift quasars, 
ULASJ112010+641 with the mass of $m_\mathrm{BH} = 2 \times 10^9 ~M_\odot$ at redshift $z=7.085$  \citep{mor11}
and SDSS J01001+2802 with $m_\mathrm{BH} = 1.2 \times 10^{10} ~M_\odot$ at $z=6.30$ \citep{wu15},
grow via mass accretion from such stellar-mass BHs, 
the Eddington ratio ($\lambda$) 
is required to be $\lambda =1.4$ for ULASJ112010+641, or $\lambda =1.3$ for SDSS J01001+2802.
However, the continuous accretion is unlikely to be sustained due to feedback, and thus 
the average mass accretion rates should be lower than the Eddington rate \citep*{alv09,mil09},
although it is pointed out that the super-Eddington accretion may be allowed in metal free systems \citep{vol05}.
The maximum rate of the super-Eddington accretion is thought to be given by Hoyle-Lyttleton accretion \citep*{hoy39,bon44}. 
However, it is not elucidated how high accretion rate is realized in an early universe. 

On the other hand, 
according to the hierarchical merger history of galaxy formation, galaxies with multiple MBHs are likely 
to form, if the number of BHs is conserved during the galaxy merger. 
However, most of galaxies harbor just one SMBH in their centres, with several
exceptions like a triple AGN in the galaxy SDSS J1027+1749 at $z = 0.066$ \citep{Liu11}, three rapidly growing
MBHs of $10^6-10^7 ~M_\odot$ in a clumpy galaxy at $z = 1.35$
\citep{Schawinski11}, a quasar triplet QQQJ1432--0106 at $z =2.076$ \citep{Djorgovski07}, and 
a second quasar triplet QQQJ1519+0627 at $z=1.51$ \citep{Farina13}. Thus, it is conceivable that
the merger of BHs might take place in some redshift epoch, resultantly forming one central SMBH. 

The merger of massive BHs may be a potential source of gravitational waves for 
the Laser Interferometer Space Antenna (LISA) and pulsar timing \citep*{Berti06, Sesana05, WL03}. 
Also, the inspiral, merger, and ringdown of binary stellar-mass black hole systems can be demonstrated
by gravitational waves with the Laser Interferometer Gravitational-Wave Observatory (LIGO),
the VIRGO \citep*{abbott06, aasi13}, the EGO 600 \citep{Huck06}, or the KAGRA \citep{Aso13}.
Very recently, the first example of gravitational waves from the merger of a $36M_\odot$ and  $29M_\odot$ BH
binary has been detected in the LIGO
\citep{abbott16}. 
These observations show that binary stellar-mass black hole systems do exist
and they can merge due to gravitational wave radiation within the cosmic time. 
Also, a weak hard X-ray transient source was detected at 0.4 s after the GW event
with {\it Fermi} Gamma-ray Burst Monitor \citep{Connaughton16}.

Recent radiation hydrodynamic simulations on the formation of first stars show 
that multiple massive stars form in a primordial gas cloud 
of $\sim 10^4-10^5 ~{M_\odot}$ with the density of around 
$10^7 ~\mathrm{cm^{-3}}$ and the extension of $\sim$0.01 pc, 
where the gas fraction is $99\%$ \citep*{gre11,ume12,sus13,sus14}. 
According to the mass function of first stars, 
multiple BHs of several tens $M_\odot$ may be born as remnants of supernovae, in such a primordial could.
In this circumstance, high mass-accretion rates onto BHs are expected. 
On the other hand,  plenty of gas can exert dynamical friction on moving BHs.
Recently, \citet{tag15} has explored the early merger of BHs through the gas dynamical friction, and
have shown that the merger time of multiple BHs merger in the gas number density of $n_\mathrm{gas}\gtrsim10^6~\mathrm{cm^{-3}}$ 
is $\sim 10^7$ yr, which is shorter than the Eddington timescale. 
However, \citet{tag15} did not consider the effect of the mass accretion onto BHs. 
Thus, in the competition between the mass accretion and the merger, 
which mechanism dominates the growth of massive BHs is not clear.

In this paper, we simulate a multiple BH system, including both the gas dynamical friction and
the gas accretion, and derive the critical condition that bifurcates the key mechanism of the growth of BHs.
In Section 2, we describe the method for numerical simulations.
In Section 3, numerical results on merger mechanisms are presented.
In Section 4, we derive the critical condition that determines a predominant merger 
mechanism, and discuss related issues, including the GW150914 event. 
Section 5 is devoted to conclusions. 

\section{Method for Numerical Simulations}

\subsection{Numerical scheme}

\citet{tag15} gives a detailed description of the simulation used here. 
We summarize some important treatments in our simulations. 

The equations of motion for BHs are given by
\begin{eqnarray}
  \frac{d^2{\bf r}_i}{dt^2} = \sum_{j}^{N_{\rm BH}} 
 \left\{ - Gm_{j}\frac{{\bf r}_{i}-{\bf r}_{j}}{|{\bf r}_{i}-
 {\bf r}_{j}|^3} + {\bf a}_{{\rm PN},ij} \right\} \nonumber \\
 + {{\bf a}_{\mathrm{acc},i}} 
+ {{\bf a}^{\rm gas}_{\mathrm{DF},i}} 
+ {\bf a}_{\mathrm{pot},i},
\end{eqnarray}
where ${\bf r}_i$ and ${\bf r}_j$ are respectively the positions of $i$-th BH and $j$-th BH, 
$N_{\rm BH}$ is the number of BHs, 
$G$ is the gravitational constant, $m_j$ is the mass  of $j$-th BH, 
${\bf a}_{{\rm PN},ij}$ is the general relativistic acceleration of $j$-th BH on $i$-th BH 
in the post-Newtonian prescription up to 2.5PN term \citep{kup06}. 
1PN and 2PN terms correspond to the pericenter shift, and 2.5PN term does to 
the gravitational wave (GW) emission. 
${{\bf a}_{\mathrm{acc},i}}$ is the acceleration of $i$-th BH due to the gas accretion, 
${{\bf a}^{\rm gas}_{\mathrm{DF},i}}$ is the acceleration of the dynamical friction (DF) on $i$-th BH by gas, 
and  ${\bf a}_{\mathrm{pot},i}$ is the acceleration on $i$-th BH by gravitational potential of gas. 

For the gas dynamical friction force, we use the formula given by \cite{tan09} 
for the motion with ${\cal M}_i<{\cal M}_\mathrm{eq}$ and  \cite{ost99} 
for ${\cal M}_i>{\cal M}_\mathrm{eq}$, where ${\cal M}_i$ is the Mach number of $i$-th BH and ${\cal M}_\mathrm{eq}$ is 
the Mach number where these two formulas give equal acceleration. 
Here, we adopt ${\cal M}_\mathrm{eq}=1.5$ as \cite{tan09}. 
Then, the acceleration of the gas dynamical friction (${{\bf a}^{\rm gas}_{\mathrm{DF},i}}$) is given by 
\begin{equation}
	{{\bf a}^{\rm gas}_{\mathrm{DF},i}} = - 4{\pi} {G}^2 m_i m_\mathrm{H}{n_\mathrm{gas}(r)}\frac{{\bf v}_i}{{v_i}^3} {\times} f({\cal M}_i) 
\end{equation}
\begin{equation}
f({\cal M}_i) =
\left \{
\begin{array}{ll}
	0.5 \mathrm{ln} \left( \frac{v_it}{{r}_\mathrm{min}} \right) \left[ \mathrm{erf} \left(\frac{{\cal M}_i}{\sqrt{2}} \right) - {\sqrt{\frac{2}{\pi}}} {\cal M}_i \mathrm{exp}( - \frac{{\cal M}_i ^2}{2}) \right]  ,\\\\
( 0 {\leq} {\cal M}_i {\leq} 0.8 ) \\\\
1.5 \mathrm{ln} \left( \frac{v_it}{{r}_\mathrm{min}} \right) \left[ \mathrm{erf} \left( \frac{{\cal M}_i}{\sqrt{2}} \right) - {\sqrt{\frac{2}{\pi}}} {{\cal M}_i} \mathrm{exp}( - \frac{{\cal M}_i ^2}{2}) \right]   ,\\\\
( 0.8 {\leq} {\cal M}_i {\leq} {\cal M}_\mathrm{eq} ) \\\\
	\frac{1}{2} \mathrm{ln} \left( 1 - \frac{1}{{\cal M}_i^2} \right) + \mathrm{ln} \left( \frac{v_it}{{r}_\mathrm{min}} \right)  ,\\\\
({\cal M}_\mathrm{eq} {\leq} {\cal M}_i )
\end{array}
\right. 
\end{equation}
where $m_\mathrm{H}$ is the mass of the hydrogen atom, $n_\mathrm{gas}$ is the number density of gas, 
$v_i$ is the velocity of $i$-th BH, and $t$ is the elapsed time. 
The $r_\mathrm{min}$ is the minimum scale of the dynamical friction on a BH, and we give $r_\mathrm{min}$ 
as $Gm_i/v_i^2$. Here, $v_it$ means the effective scale of gas medium, and we set an upper 
limit of $v_it$ to $0.1 ~\mathrm{pc}$.  
When $v_it < r_\mathrm{min}$, we assume $f({\cal M}_i)=0$.

The equations of motion is integrated using the fourth-order Hermite scheme with the shared time step \citep{mak92} 
whose accuracy parameter is set to be $0.003$. 
To use the fourth-order Hermite scheme, we calculate the time derivative of the acceleration by the Newtonian gravity, 
the gas gravitational potential, and the relativistic force. 
On the other hand, we treat the dynamical friction and the accretion of gas to quadratic order.

\subsection{Mass accretion rate}\label{mass-accretion-rate}

We envisage the Hoyle-Lyttleton accretion onto BHs. 
The mass accretion rate is not limited by the Eddington luminosity in a BH accretion disk,
since photons are trapped in innermost optically-thick regions without diffusing out
from the disk surface (so-called photon trapping effects) \citep{Abramowicz88}. 
Here, we employ a model for such a super-Eddington accretion.  
The luminosity of a super-Eddington disk is fitted as
\begin{eqnarray}
	\Gamma& \equiv &L/L_\mathrm{E} \nonumber\\
			&\simeq&
\left \{
\begin{array}{ll}
	2\left[ 1+\mathrm{ln}\left ( \frac{\dot{m}}{20\eta\dot{m}_\mathrm{E}}\right ) \right] ~~\mathrm{for}~\dot{m} \geq 20~\eta\dot{m}_\mathrm{E}  \\
	\left ( \frac{\dot{m}}{10\eta\dot{m}_\mathrm{E}}\right )~~~~~~~~~~~~~ \mathrm{for}~\dot{m} < 20~\eta\dot{m}_\mathrm{E} 
\end{array}
\right. 
\label{eq_wat}
\end{eqnarray}
for the viscosity parameter $\alpha=0.1$ \citep{wat00}, where 
$L_\mathrm{E}$ is the Eddington luminosity, $\dot{m}$ is the mass accretion rate, 
$\dot{m}_\mathrm{E}$ ($=L_\mathrm{E}/\eta c^2$) is the Eddington accretion rate, 
$\eta$ is the radiative energy conversion efficiency ($\eta=0.1$ in this paper),
and $c$ is the light speed. 

\begin{figure}
\centering
\includegraphics[width=85mm]{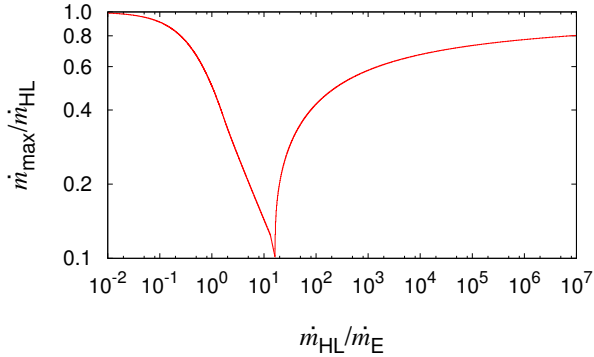}
\vspace{-5mm}
 \caption{The maximum gas accretion rate ($\dot{m}_{\rm max}$) 
with the reduction by radiation pressure
as a function of the Eddington accretion rate ($\dot{m}_{\rm E}$), where
$\dot{m}_\mathrm{HL}$ is the Hoyle-Lyttleton accretion rate.
A viscous accretion disk with viscosity parameter $\alpha=0.1$ is assumed.
}
\label{hlac_rad}
\end{figure}

Also, we take the effect of the radiation pressure on the Hoyle-Lyttleton accretion
into consideration. 
The accretion rate is reduced due to the radiation pressure as
\begin{eqnarray}
	\frac{\dot{m}}{\dot{m}_\mathrm{HL}}
	= 
\left \{
\begin{array}{ll}
	(1-\Gamma) ~~~~~~~~~~~~~~~~~(0\leq \Gamma \leq 0.64 ) \\
	(1-\Gamma-\frac{2}{\pi}\psi_0+\Gamma \mathrm{sin}\psi_0)\\
	~~~~~~~~~~~~~~~~~~~~~~~~~~( 0.64\leq \Gamma \leq 1.65 ) \\
	\frac{2}{\pi}\mathrm{tan}^{-1}\left ( \frac{\Gamma}{0.75\mathrm{ln}10^5}\right ) ~~( 1.65\leq \Gamma )  
\end{array}
\right. 
\label{eq_han}
\end{eqnarray}
\citep{han01}, where 
$\dot{m}_\mathrm{HL}$ is the Hoyle-Lyttleton accretion rate,
and $\mathrm{cos}\, \psi_0=2/(\pi\Gamma)$. 
By combining equations (\ref{eq_wat}) and (\ref{eq_han}), we can derive 
the maximum gas accretion rate ($\dot{m}_{\rm max}$),
which is shown in Fig. \ref{hlac_rad}. 
This figure shows that the Hoyle-Lyttleton accretion rate is maximally reduced 
down to $\sim 0.1~\dot{m}_\mathrm{HL}$ around 
$\dot{m}_{\rm E}/\dot{m}_\mathrm{HL} \approx 
10
$.
For $\dot{m}_{\rm E}/\dot{m}_\mathrm{HL} > 
10
$, 
the reduction is alleviated owing to the photon trapping effect. 

To parameterize the accretion rate, we set the gas mass accretion rate as 
\begin{equation}
	\dot{m}_{i}=\epsilon \dot{m}_{\mathrm{HL},i}=\epsilon \frac{4\pi G^2 m_\mathrm{H} n_\mathrm{gas}m_{i}^2}{(C_\mathrm{s}^2+v_i^2)^{3/2}}, 
\label{eq_rhoyle}
\end{equation}
where $v_i$ is the velocity of $i$th BH, and $\epsilon (\leq 1)$ is the accretion efficiency. 
In this paper, taking another feedback effects into consideration, we consider the range of 
$10^{-7} \leq \epsilon \leq 1$. 
For a given $\dot{m}_{i}$, it is limited to the maximum accretion rate.
As for the back reaction by gas accretion, 
we incorporate the acceleration due to mass accretion. 
If we assume the relative velocity between gas and BH to be 
the BH velocity  (i.e., static ambient gas on the Hoyle-Lyttleton accretion scales), 
the acceleration due to mass accretion on $i$th BH $a_{\mathrm{acc},i}$ is given as
\begin{equation}
	a_{\mathrm{acc},i}=-\frac{\dot{m}_iv_i}{m_i}.
\label{eq_acc}
\end{equation}
Also, when the gas accretes onto BHs, we lessen the gas mass and correspondingly 
the gas number density to conserve the total mass of the system. 

\subsection{Setup of Simulations}

One key parameter in our simulations is the BH density, $\rho_\mathrm{BH}$, at the initial epoch. 
Recent radiation hydrodynamic simulations on the formation of the first stars have shown that 
several or more stars are born in a disk of $\sim 0.01 ~\mathrm{pc}$ \citep*{gre11,ume12,sus13,sus14}. 
Referring to these results, we change the typical extensions of BH distributions 
at the initial epoch, $r_\mathrm{typ}$, to settle the BH density. 
We investigate the range from  $0.01 ~\mathrm{pc}$ to $0.1 ~\mathrm{pc}$ in this paper.
As first star remnants, we set up ten BHs of equal mass of $30~M_\odot$ initially. 

Another key parameter is the gas number density $n_\mathrm{gas}$. 
Simulations of first star formation show that $n_\mathrm{gas}$ is $\sim 10^{7-8} ~\mathrm{cm^{-3}}$ 
in a primordial cloud of $\sim 0.01 ~\mathrm{pc}$. 
In this paper, we consider a wider range of the gas density
to explore the possibility of BH mergers in a wide variety of environments. 
In dense interstellar cloud cores, $n_{\rm gas}$ ranges from $10^{5}{\rm cm}^{-3}$ to $10^{7}~{\rm cm}^{-3}$
\citep[e.g.][]{Bergin96}.
In galactic nuclear regions, $n_{\rm gas} \gtrsim 10^{8}~{\rm cm}^{-3}$ at $\lesssim 1{\rm pc}$ \citep[e.g.][]{NU16}.
Anyhow the gas density is less than $10^{10}~\mathrm{cm}^{-3}$ in realistic environments, but
we consider even higher densities up to $10^{12}~\mathrm{cm}^{-3}$ as well in order to elucidate more clearly 
the dependence of merger criteria on the gas density. 

\begin{figure*}
\begin{center}
\includegraphics[width=170mm]{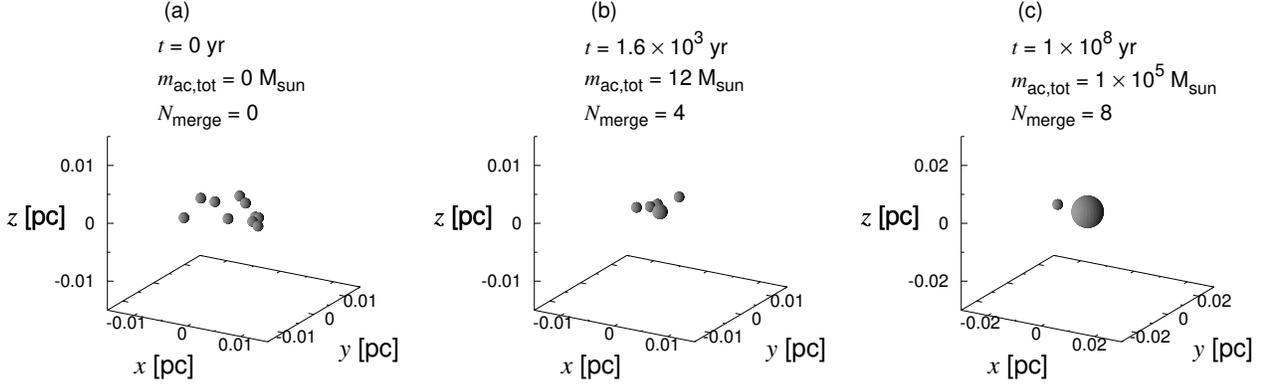}
\caption{
	Snapshots of the distributions of the multiple BHs for a gas drag-driven merger (type A). 
	The initial parameters are a typical extension of  BH distributions, $r_\mathrm{typ}=0.01~\mathrm{pc}$, 
	the gas number density, $n_\mathrm{gas}=10^{12}~\mathrm{cm^{-3}}$, and the accretion 
	efficiency, $\epsilon=10^{-4}$.
	In each snapshot, the elapsed time $t$, the total accreted mass $m_\mathrm{ac,tot}$ onto BHs, 
and the number of merged BHs $N_\mathrm{merge}$ are presented. 
	The sizes of spheres represent the mass of the BH in logarithmic scales, 
	where the smallest one corresponds to the initial BH mass, $m_i = 30 M_\odot$. 
}
	\label{3d12}
\end{center}
\end{figure*}

\begin{figure}
\vspace{-5mm}
\hspace*{-8mm}
\includegraphics[width=95mm]{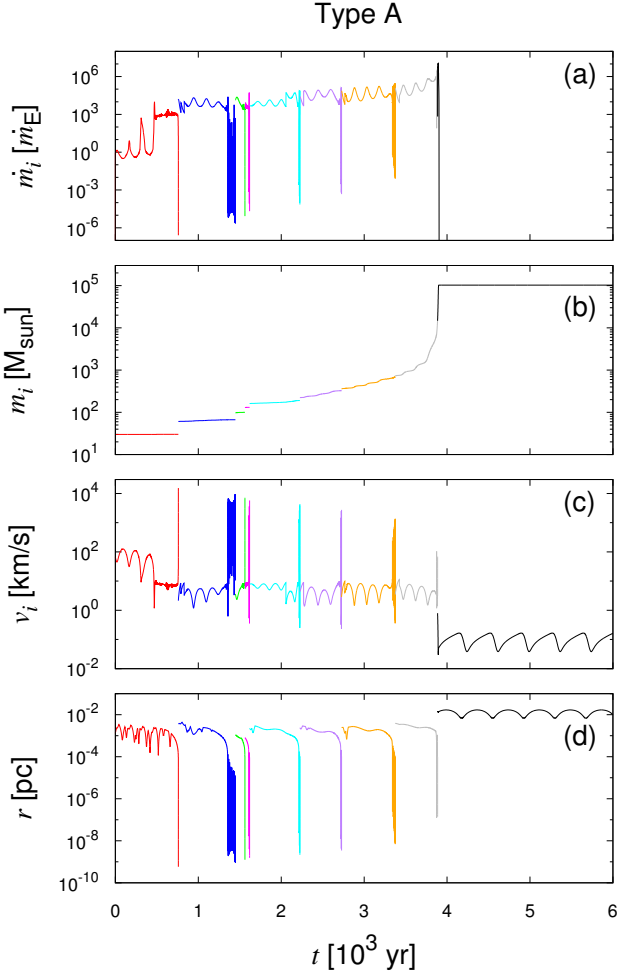}
\vspace{-8mm}
\caption{
	The time evolution of the physical quantities for a gas drag-driven merger (type A). 
	The initial parameters are the same as Figure \ref{3d12}.
	Panels (a), (b), and (c) represent the mass accretion rate in units of
the Eddington mass accretion rate for $30 M_\odot$, 
the mass and the velocity of the first merged BH,  respectively. 
	Panel (d) represents the separation of the closest pair within all BHs,  
	where the colors of lines change at every event of the BH merger. 
}
\vspace{0mm}
	\label{vt12}
\end{figure}

Although \citet{tag15} assumed that gas is distributed infinitely, 
we settle the finite gas mass in a sphere with $M_\mathrm{gas, tot}=10^5~{M_\odot}$ in this paper. 
Outside this sphere, BHs are not subject to gas dynamical friction and gas accretion. 
In realistic environments, the density distributions of gas are likely to be fairly complex.
In first-generation objects, the axis ratios of gas disks ranges from one to five \citep{Hirano14}, and also 
the gas flow by cold accretion along filaments may change the gas distribution significantly 
at high redshifts \citep[e.g.][]{Yajima15}. 
In dense molecular cloud cores, the gas distributions are sometimes fairly spherical and sometimes elongated.
Since the gas distributions depend on objects of interest, we assume a spherical cloud
in this paper just for simplicity. 
The temperature of the gas is set to be $1000 ~\mathrm{K}$, based on the thermal history of the metal poor gas \citep{omu00}. 
Consequently, the sound speed is given as $C_\mathrm{s}=3.709 ~\mathrm{[km/s]}$. 
Furthermore, we change the gas accretion efficiency, $\epsilon$, as an important parameter,
which is constrained as described in the previous section. 
Since gas accretion rate in a first-generation object is not elucidated, 
we study a wider range of gas accretion rate as $\epsilon$ from $10^{-7}$ to $1$. 

We assume that two MBHs merge, when their separation is
less than 100 times the sum of their Schwarzschild radii:
\begin{equation}
  \left|\bm{r}_{i} - \bm{r}_{j} \right| < 100
  \left( r_{{\rm sch},i} + r_{{\rm sch},j} \right),
\label{merge_con}
\end{equation}
where $r_{{\rm sch},i}$ is the Schwarzschild radius of $i$-th BH
given by $2Gm_{i}/c^2$ with the speed of light $c$.
Therefore, the simulations do resolve the scales of $100~r_{{\rm sch}}=2 \times 10^9~{\rm cm}$.
To avoid cancellation of significant digits when such tiny scales as $100~r_{{\rm sch}}$ are resolved, 
the BHs evolution is calculated in the coordinate, 
where the origin is always set to the center of mass for the closest pair of BHs. 
This prescription allows us to pursue accurately the orbit of the BHs until the merger condition is satisfied. 
At the final stage of merger, the binding energy of a binary BH is transformed to the energy of GW, 
retaining the mass of each BH.

We give the initial positions of BHs randomly in the $x-y$ plane within $r_\mathrm{typ}$. 
Also, the velocity of each BH is given as the sum of a circular component and a random component. 
The circular velocity is given to balance against the gravity of gas in the $x-y$ plane. 
Besides, we give the random velocity in the $xyz$ space 
according to a Gaussian distribution with the same dispersion as the circular velocity. 

Our simulations are performed for $100~\mathrm{Myr}$, since the background environments 
of the host objects are likely to change in $100~\mathrm{Myr}$. 
Also, we terminate the simulation, if all BHs merge into one BH, or if all other BHs are escaped except one BH.

\begin{figure*}
\includegraphics[width=170mm]{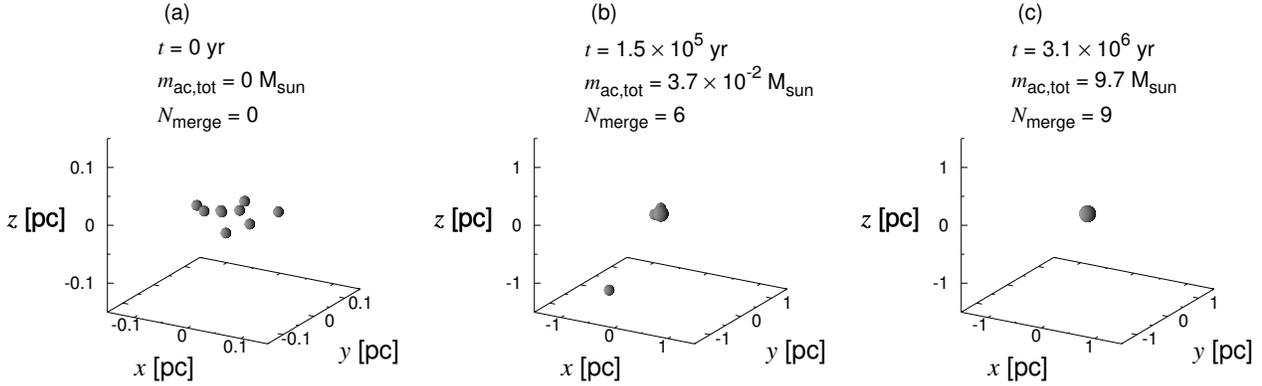}
\caption{
	Same as Fig. \ref{3d12}, but for an interplay-driven merger (type B). 
$r_\mathrm{typ}=0.1~\mathrm{pc}$, $n_\mathrm{gas}=10^{9}~\mathrm{cm^{-3}}$, and $\epsilon=10^{-6}$. 
}
\label{3d87}
\end{figure*}

\begin{figure}
	\vspace{-5mm}
\hspace*{-8mm}
\includegraphics[width=95mm]{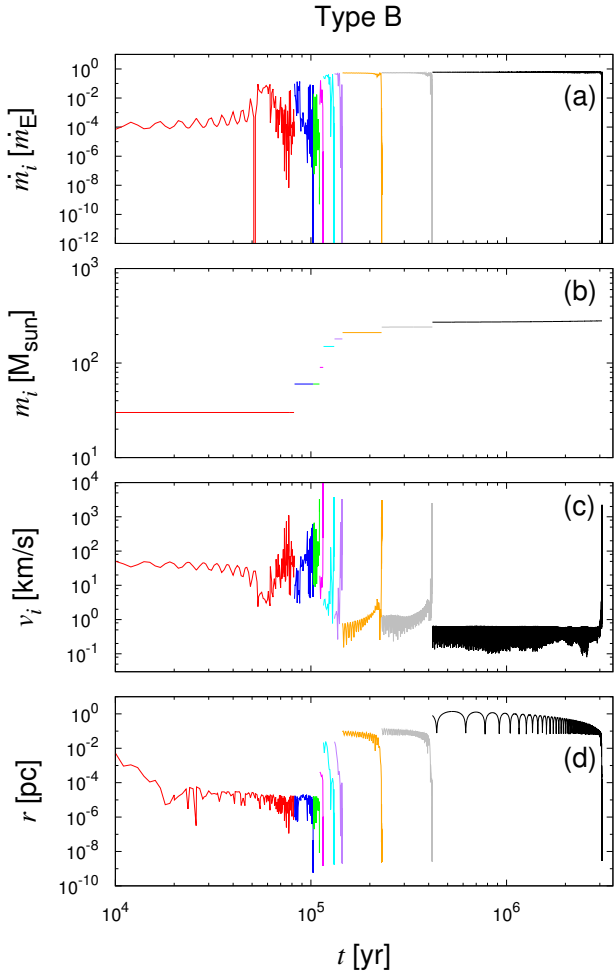}
\vspace{-8mm}
\caption{
	Same as Fig. \ref{vt12}, but for an interplay-driven merger (type B). 
	$r_\mathrm{typ}=0.1~\mathrm{pc}$, $n_\mathrm{gas}=10^{9}~\mathrm{cm^{-3}}$, and $\epsilon=10^{-6}$. 
}
\label{vt87}
\end{figure}

\section{Merger mechanisms}

Here, we present the merger mechanisms depending upon the gas density and accretion rate.
In our previous study without mass accretion \citep{tag15}, the merger mechanisms are categorized into three types: 
a gas drag-driven merger (type A), an interplay-driven merger (type B), and a three body-driven merger (type C). 
In this study, we find an additional merger mechanism, which is referred to as an accretion-driven merger (type D). 
The classification is based on 
the manner of orbit decay just before the gravitational wave emission drives the merger. 
In type A, the orbit is decayed due to the dynamical friction by gas before the gravitational wave works. 
In both of types B and C, the strong disturbance of the orbit is induced by three-body interaction during the first merger. 
But in type B, after first few mergers that are promoted by three-body interactions,
the separations of BHs are increased due to the slingshot mechanism. Thereafter,
the orbits of BHs decay slowly for long time through the gas dynamical friction. 
Whether the slow mergers occur from increased separations is the criterion on which
type B is discriminated from type C. 
In type C, the strong disturbance of the orbit due to three-body interactions continues until the final merger. 
In type D, significant accretion occurs before the first merger. 
In the following, we scrutinize the effects of gas accretion on each merger mechanism.

\subsection{Gas drag-driven merger (type A)}

In Fig. \ref{3d12}, the snapshots of the distributions of multiple BHs 
for a gas drag-driven merger (type A) are shown in three different stages,
where a typical extension of  BH distributions is initially $r_\mathrm{typ}=0.01~\mathrm{pc}$, 
the background gas number density is as high as $n_\mathrm{gas}=10^{12}~\mathrm{cm^{-3}}$, and 
the accretion efficiency is $\epsilon=10^{-4}$.
The sizes of spheres represent the mass of the BH in logarithmic scales, 
where the smallest one corresponds to the initial BH mass, $m_i = 30 M_\odot$. 
In each snapshot, the elapsed time $t$, the total accreted mass $m_\mathrm{ac,tot}$ onto BHs, 
and the number of merged BHs $N_\mathrm{merge}$ are presented. 
As shown in this figure, several BHs merge without considerable mass accretion before $1.6 \times 10^3{\rm yr}$.
Thereafter, subsequent mergers proceed and eventually all gas accretes onto the most massive BH
before $10^8{\rm yr}$.

\begin{figure*}
	\begin{center}
\includegraphics[width=170mm]{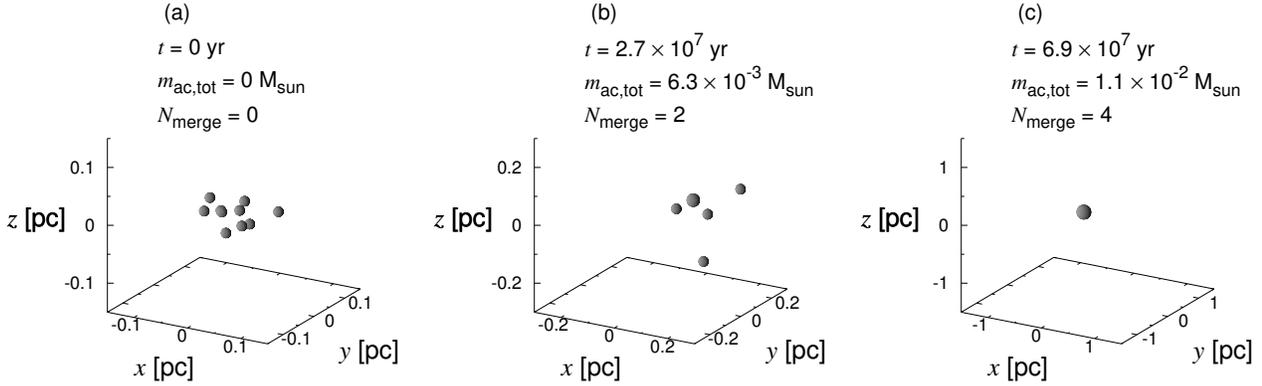}
\caption{
	Same as Fig. \ref{3d12}, but for a three-body-driven merger (type C). 
$r_\mathrm{typ}=0.1~\mathrm{pc}$, $n_\mathrm{gas}=10^{6}~\mathrm{cm^{-3}}$, and $\epsilon=10^{-6}$. 
}
\label{3d9}
\end{center}
\end{figure*}
\begin{figure}
	\vspace{-5mm}
\hspace*{-8mm}
\includegraphics[width=95mm]{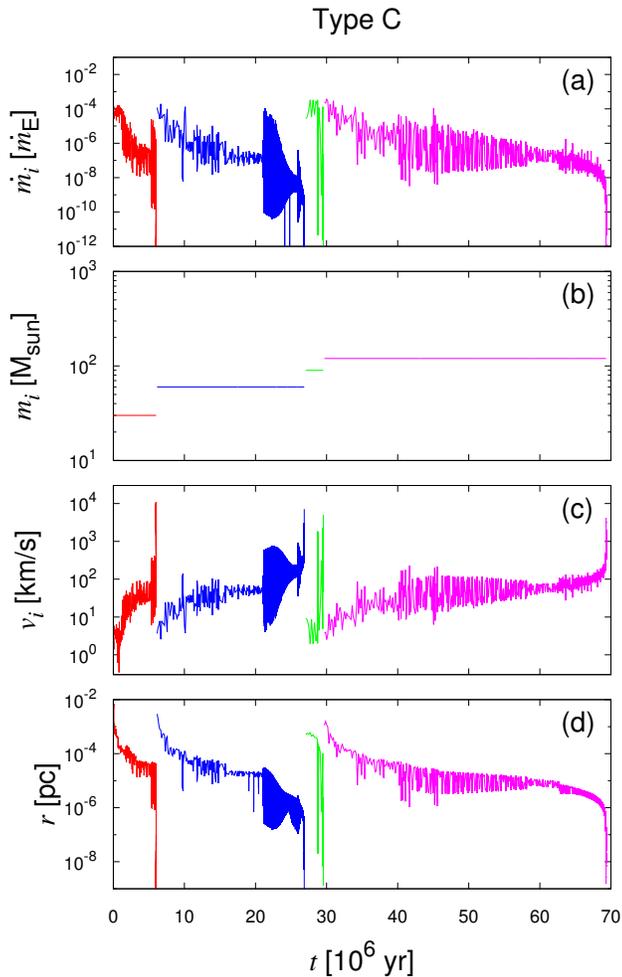}
\vspace{-8mm}
\caption{
	Same as Fig. \ref{vt12}, but for a three-body-driven merger (type C). 
	$r_\mathrm{typ}=0.1~\mathrm{pc}$, $n_\mathrm{gas}=10^{6}~\mathrm{cm^{-3}}$, and $\epsilon=10^{-6}$. 
}
\label{vt9}
\end{figure}

To see the detailed physical processes, we present, in Fig. \ref{vt12}, the time evolution of the mass accretion rate, 
the mass and the velocity of the first merged BH,  and the separation of the closest pair within all BHs. 
In a type A merger, the separation of BHs decays effectively due to the dynamical friction (DF) . 
The DF timescale is given by 
\begin{equation}
	t_\mathrm{DF}\simeq \frac{v^3}{4\pi G^2 m_\mathrm{BH}m_\mathrm{H} n_\mathrm{gas}}.
\end{equation}
On the other hand, the accretion timescale, at which the accretion rate diverges, is given as 
\begin{equation}
	t_\mathrm{ac}=\frac{(v^2+C_\mathrm{s}^2)^{3/2}}{4\pi G^2 m_\mathrm{BH}m_\mathrm{H} n_\mathrm{gas}\epsilon}.
\end{equation}
It is noted that the accretion timescale has the same dependence as the DF timescale 
on the BH mass and the gas density, 
because the Hoyle-Lyttleton accretion and the DF are both caused by the change of the streamline 
due to the gravity of a BH. 
However, the velocity-dependence is different from each other. 
The accretion timescale is longer than the DF timescale, since 
the accretion timescale depends on the sound velocity and also 
on the accretion efficiency $\epsilon(\leq1)$. 
But, as in the first merger shown in Fig.3,  
the accretion timescale is regardless of the sound speed, if  the velocity is supersonic (panel (c)).
On the other hand, the accretion timescale is increased by a low accretion efficiency as $\epsilon=10^{-4}$. 
Therefore, the decay of orbit is predominantly driven by the gas dynamical friction (panel (d)) 
before significant gas accretion. 
As a result, a close BH binary forms and eventually merges into one BH due to the gravitational wave radiation before $10^3$ yr.
Thereafter, although there is a long phase when the BH velocity is low and the accretion rate is accordingly raised, 
the DF timescale is still shorter than the accretion timescale. 
Hence, successive mergers proceed by the dynamical friction. 
It is worth noting that the velocity is highly supersonic at the moment of merger owing to the inspiral
by friction and therefore the accretion rate is much lower than the Eddington rate.
Finally, when all gas instantly accretes onto the most massive BH, 
the merger stops and a binary BH is left in the system as seen in panel (c) of Fig. \ref{3d12}. 

\begin{figure*}
	\begin{center}	
\includegraphics[width=170mm]{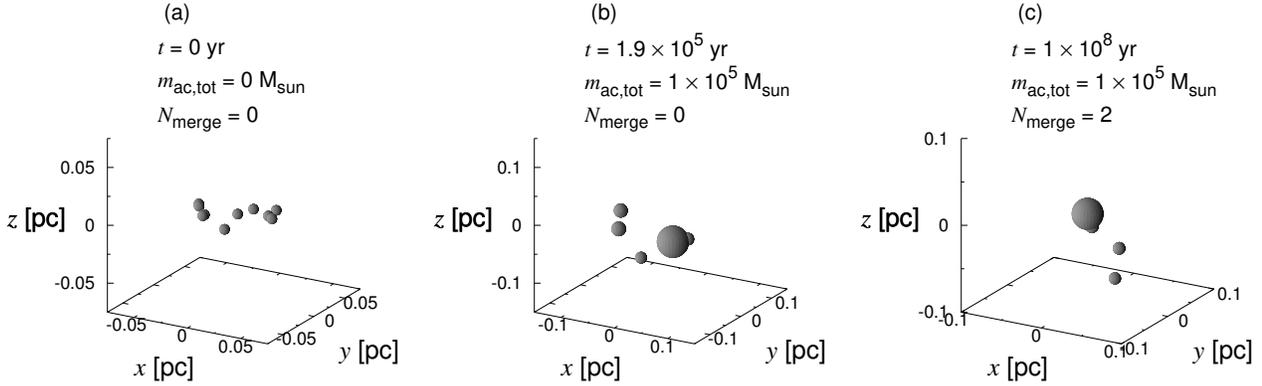}
\caption{
	Same as Fig. \ref{3d12}, but for an accretion-driven merger (type D). 
$r_\mathrm{typ}=0.04~\mathrm{pc}$, $n_\mathrm{gas}=10^{7}~\mathrm{cm^{-3}}$, and $\epsilon=10^{-1}$. 
}
\label{3d46}
\end{center}
\end{figure*}

\begin{figure}
	\vspace{-5mm}
	\hspace*{-8mm}
\includegraphics[width=95mm]{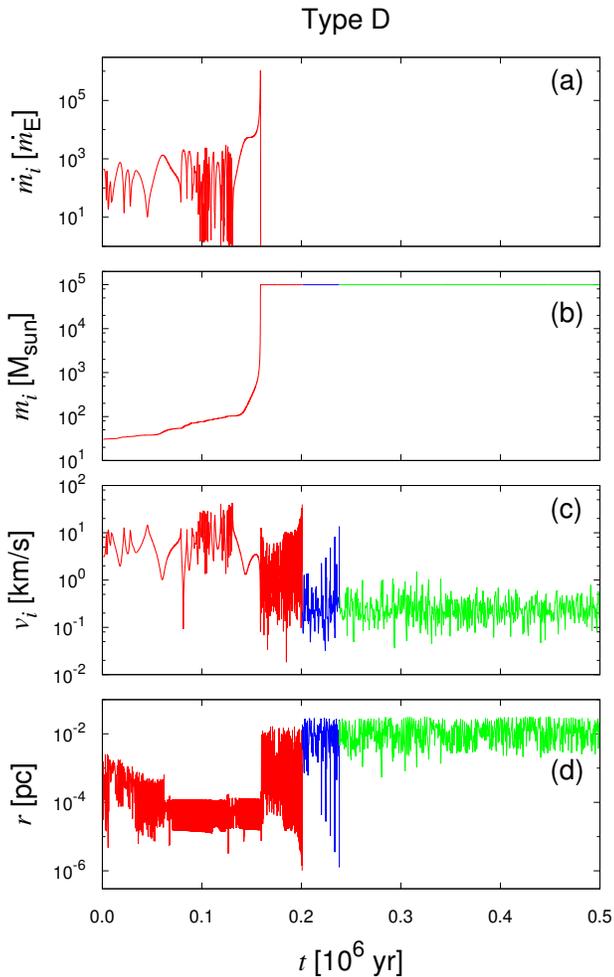}
\vspace{-8mm}
\caption{
	Same as Fig. \ref{vt12}, but for an accretion-driven merger (type D). 
	$r_\mathrm{typ}=0.04~\mathrm{pc}$, $n_\mathrm{gas}=10^{7}~\mathrm{cm^{-3}}$, and $\epsilon=10^{-1}$. 
}
\label{vt46}
\end{figure}

\subsection{Interplay-driven merger (type B)}

An example of an interplay-driven merger (type B) is presented in Fig. \ref{3d87} and Fig. \ref{vt87}.
In this example, the accretion efficiency is as low as $\epsilon=10^{-6}$.
As seen in panel (d) of Fig. \ref{vt87}, strong irregular oscillations of the separation of BHs occur in the first several mergers. 
If they are simply the Keplerian orbital motions of a binary,
both of the pericenter and apocenter shrink smoothly due to the gas drag. 
However, in Fig 5, the pericenter and apocenter change promptly many times in the first few mergers. 
Such discontinuous changes of the pericenter and apocenter are typical in the three-body encounters. 
Hence, the successive mergers are thought to be promoted by the three-body interaction. 
On the other hand, in the last few mergers, 
the orbital evolution starts from a larger separation than the initial typical separation 
and the separation of BHs decays smoothly due to the dynamical friction. 
This increase of separation is a negative effect of three-body interaction. 
The panel (b) of Fig. \ref{3d87} shows that a BH is kicked out due to slingshot mechanism. 
As a result of this negative effect, the velocity becomes much lower than the sound velocity 
due to the deceleration by the gas potential (see panel (c) in Fig. \ref{vt87}),
so that the accretion rate is raised in the later phase. 
However, mergers by the dynamical friction proceed faster than the accretion, because 
the accretion efficiency is quite low compared to the Hoyle-Lyttleton accretion. 
Eventually, all BHs merge into one within $3 \times 10^6{\rm yr}$, and all the accreted mass
is $9.7 M_\odot$.

\subsection{Three body-driven merger (type C)}

In a three body-driven merger (type C), the strong disturbance of the orbit continues until the final merger.
The three-body interaction of BHs solely transfers the angular momentum, eventually
causing the merger via the gravitational wave radiation.
An example of type C is presented in Fig. \ref{3d9} and Fig. \ref{vt9}. 
The initial parameters are the same as those in the above type B merger, excepting for
lower background gas density. 
Strong oscillations of separation of BHs seen in panels (c) and (d) of Fig. \ref{vt9} 
are caused by the three-body encounters. 
In a type C, since each BH of a closest pair receives the gravity mainly from another BH, 
the velocities of BHs in a closest pair increase according to the decay of the separation through
three-body interactions, keeping supersonic velocities. 
Therefore, the accretion rates of the BHs of closest pair for type C are much lower than those in a type B. 
On the other hand, several BHs are kicked out from the central regions 
due to the slingshot mechanism before the first few mergers,
and eventually a few BHs escape from the system. 
These escaping BHs are decelerated by the gas gravity, and therefore the velocities become as low as the sound velocity. 
The accretion rates onto these escaping BHs are often higher than those of the BHs of closest pair, but
they never join the mergers of BHs. 
As a result, only multiple BHs left in the system  merge into one BH.

\begin{figure*}
\hspace*{-0.5cm}
\includegraphics[width=185mm]{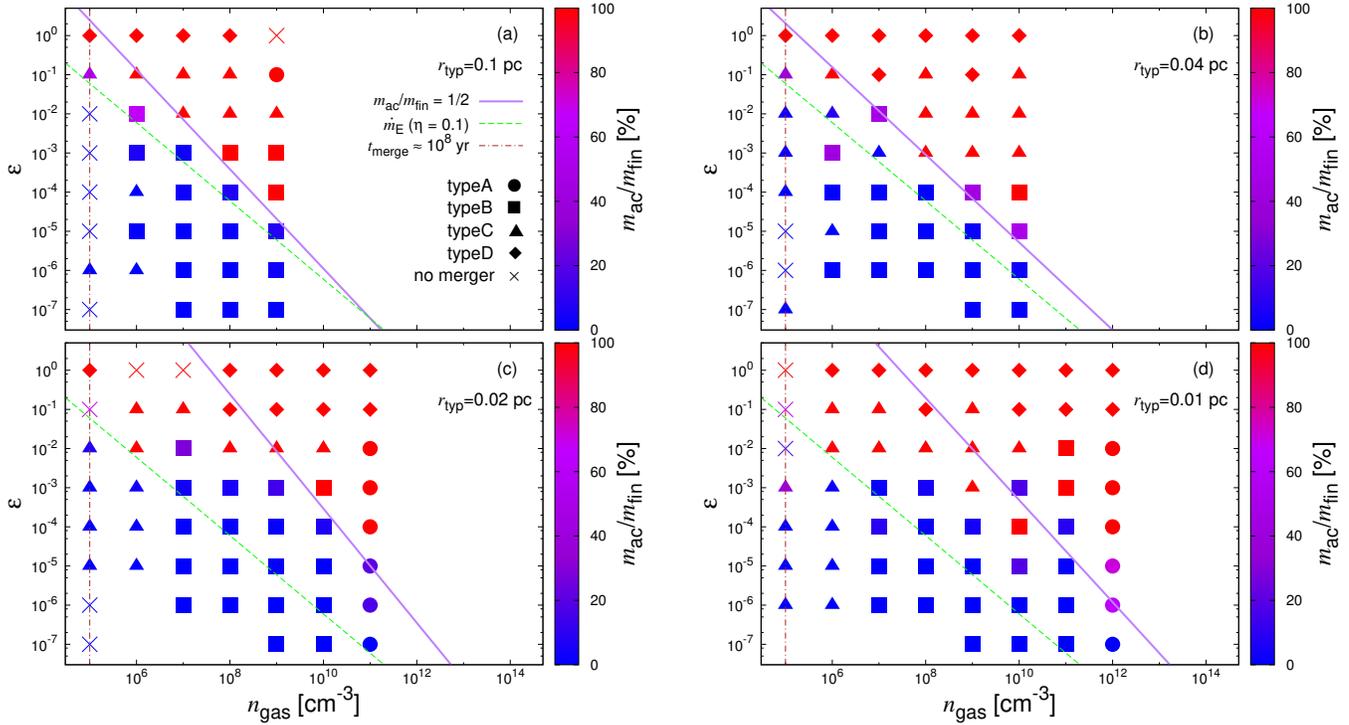}
 \caption{The contribution of the accreted mass to the final mass of the most massive BH 
	 as functions of the gas density ($n_\mathrm{gas}$) and the accretion 
	 efficiency ($\epsilon$). 
	 Filled circles, squares, triangles, diamonds, and crosses represent gas drag-driven mergers (type A), 
	 interplay-driven mergers (type B), three body-driven mergers (type C), 
	 accretion-driven mergers (type D), and no merger, respectively. 
	 The initial typical extension of BH distributions is (a) $r_\mathrm{typ}=$0.1 pc, (b) 0.04 pc, (c) 0.02 pc, and (d) 0.01 pc. 
	 The green dashed line represents the Eddington accretion rate $\dot{m}_\mathrm{E}$ for the initial mass of the BH. 
	 The purple line represents the critical condition that bifurcates the key mechanisms for the growth of the BH. 
	 The brown dot-dashed line represents the lower limit of gas density for the occurrence of merger within $10^8$ yr, which
	 is roughly estimated from Fig. 6 in \citet{tag15}. 
 }
\label{r01}
\end{figure*}

\subsection{Accretion-driven merger (type D)}

In an accretion-driven merger (type D), significant accretion occurs before the first merger.
A type D emerges with a high accretion rate near to the Hoyle-Lyttleton rate. 
Fig. \ref{vt46} shows a typical evolution in a type D. 
In this model, the mass accretion is not high on the closest pair of BHs,
because the accretion rate is reduced by their high circular velocities, according to equation (\ref{eq_rhoyle}).
Instead, an isolated BH, which has low velocity and drifts in the outer regions for long time,
grows first by accretion. In Fig. \ref{vt46}, the accretion rate, growth of mass, and velocity
of the heaviest BH are shown. 
In this model, the BH swallows almost all gas during the drift with roughly sound velocity
in $2\times 10^5{\rm yr}$. 
Thereafter, the BH interacts gravitationally with other smaller BHs.
Since the grown-up BH has a larger Schwarzschild radius, 
the merger condition (\ref{merge_con}) is liable to be satisfied. 
Shortly after the total accretion, two BHs approaching the heaviest BH merge 
by the gravitational wave radiation. 
Then, several BHs are left without merger in the system until the end of simulation. 
Hence, the merger of all BHs into one BH is hard to occur in a type D.

\section{Merger versus accretion}

\subsection{Critical condition}

A significant measure for the growth of massive BHs is provided by 
the condition under which the predominant mechanism of the BH growth 
is bifurcated between the merger and the accretion. 
In Fig. \ref{r01}, resultant merger types are summarized 
with the contribution of the accreted mass $m_\mathrm{ac}$ 
to the final mass $m_\mathrm{fin}$ at the end of simulations. 
This figure shows that the fractions of accreted mass are a steep function 
in a two parameter plane of $\epsilon$ and $n_{\rm gas}$, depending on the extension of BH distributions. 
This behavior comes from the fact that the Hoyle-Lyttleton-type accretion rate is a nonlinear function
of mass as shown in equation (\ref{eq_rhoyle})  and therefore diverges at the finite time as 
\begin{equation}
	m_{i}=\frac{1}{m_0^{-1}-\alpha t}, 
\label{eq_hoyle_mass}
\end{equation}
where $m_0$ is the initial mass and  
$\alpha=\epsilon 4\pi G^2 m_\mathrm{H} n_\mathrm{gas}/(C_\mathrm{s}^2+v_i^2)^{3/2}$.

In the environments of low gas density, if the velocity of each BH is near to $C_\mathrm{s}$,
$t_\mathrm{DF}$ and $t_\mathrm{ac}$ are not dependent on the BH density, $\rho_\mathrm{BH}$,
furthermore, both $t_\mathrm{DF}$ and $t_\mathrm{ac}$ have the same dependence on $n_\mathrm{gas}$. 
Therefore, the critical accretion efficiency, $\epsilon_\mathrm{c}$, above which
the accretion mass becomes predominant, is expected to be $\epsilon_\mathrm{c}=\mathrm{const}$,
irrespective of $\rho_\mathrm{BH}$ and $n_\mathrm{gas}$. 
Here, to assess $\epsilon_\mathrm{c}$, we employ the average of accretion mass-to-final mass ratios, 
$m_\mathrm{ac}/m_\mathrm{fin}$. 
Also, the numerical results used in this assessment are restricted to those
in low-density regions of $n_\mathrm{gas} < 10^8 {\rm cm}^{-3}$. 
Then, we derive $\epsilon_\mathrm{c}$ by linear interpolation of 
$m_\mathrm{ac}/m_\mathrm{fin}$ between 
$\max \{ \log \epsilon \}~\mathrm{for}~2m_\mathrm{ac}<m_\mathrm{fin}$ 
and $\min \{ \log \epsilon\}~\mathrm{for}~2m_\mathrm{ac}>m_\mathrm{fin}$. 
As a result, we find $\epsilon_\mathrm{c}=6\times10^{-3}$. 
This value is broadly consistent with the condition that the accretion timescale 
is a few hundred times shorter than the timescale for the merger of ten BHs,
as discussed in \citet{tag15}.

\begin{figure}
\hspace*{-0.5cm}
\includegraphics[width=90mm]{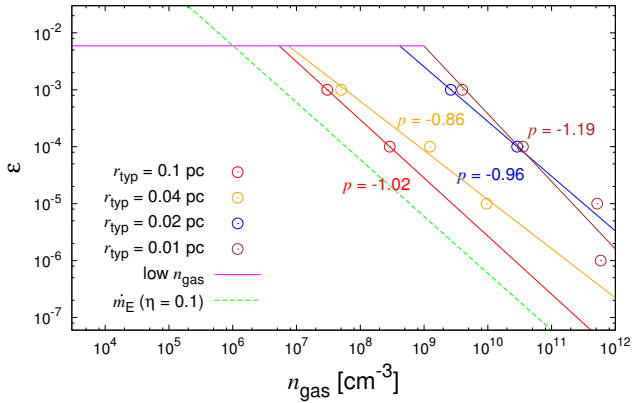}
 \caption{
	 The critical accretion efficiency, $\epsilon_\mathrm{c}$, as a function of
ambient gas density $n_\mathrm{gas}$. 
	 Red, orange, blue, and brown plots represent the critical condition in high-density regions 
	 for $r_\mathrm{typ}=0.1,~0.04,~0.02,$ and $0.01~\mathrm{pc}$, respectively. 
	 Red, orange, blue, and brown lines represent the curves fitted by 
		$n_\mathrm{gas,c}=a\epsilon^{p}$ for $r_\mathrm{typ}=0.1,~0.04,~0.02,$ and $0.01~\mathrm{pc}$. 
	 Pink line represents the critical condition in low-density regions. 
	 The green dashed line represents the Eddington accretion rate $\dot{m}_\mathrm{E}$.
 }
\label{line}
\end{figure}

\begin{figure}
\centering
\includegraphics[width=85mm]{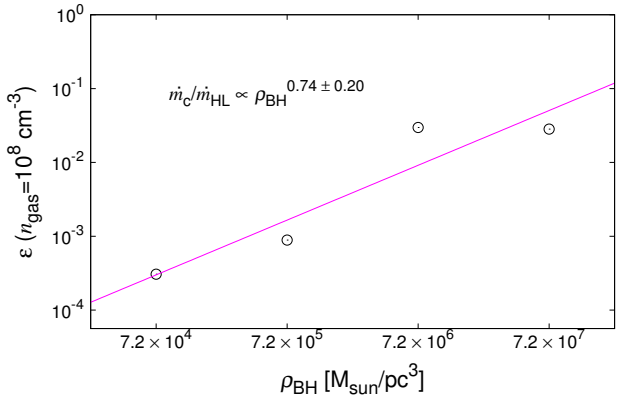}
	\caption{The critical accretion efficiency, $\epsilon_\mathrm{c}$, as a function of
$\rho_\mathrm{BH}$, for $n_\mathrm{gas}=10^8~\mathrm{cm}^{-3}$.
		A straight line is the fitting using a power law of 
		$\epsilon(n_\mathrm{gas}=10^8~\mathrm{cm}^{-3})=a \rho_\mathrm{BH}^{q}$. 
	}
	\label{beki}
\end{figure}

On the other hand, in high-density regions, 
the boundary of accretion-dominant branches is depending on $n_\mathrm{gas}$ and $r_{\rm typ}$. 
To derive the dependence of the critical accretion efficiency on $n_\mathrm{gas,c}$, 
we use the results in high-density regions of $n_\mathrm{gas} \geq 10^8 {\rm cm}^{-3}$ 
for a given BH density, say $r_\mathrm{typ}$.
In Fig. \ref{line}, we show the critical accretion efficiency, $\epsilon_\mathrm{c}$, as a function of
$n_\mathrm{gas}$. 
As shown in Fig. \ref{line}, the dependence on $n_\mathrm{gas}$ seems similar, 
almost irrespective of
$r_\mathrm{typ}$. 
Actually, if we assume a power law as $\epsilon_{\rm c}=a n_\mathrm{gas}^p$, we can fit the results
with  $p=-1.01\pm0.07$, 
by linear interpolation of $m_\mathrm{ac}/m_\mathrm{fin}$ between 
$\max \{\mathrm{log}\, n_\mathrm{gas}\}~\mathrm{for}~2m_\mathrm{ac}<m_\mathrm{fin}$ 
and $\min \{\mathrm {log} \, n_\mathrm{gas}\}~\mathrm{for}~2m_\mathrm{ac}>m_\mathrm{fin}$. 
Next, we determine the dependence on $\rho_\mathrm{BH}$.
In Fig. \ref{beki}, we show the critical accretion efficiency as a function of
$\rho_\mathrm{BH}$. The dependence seems to be well fitted by a power law form again.
If we assume $\epsilon_\mathrm{c}=a n_\mathrm{gas}^{-1.01}\rho_\mathrm{BH}^{q}$, 
we find the best fit value as $q=0.74 \pm 0.20$.

Combining the results for low and high gas density cases, the critical accretion efficiency is given as
\begin{eqnarray}
	\epsilon_\mathrm{c}=
\left \{
\begin{array}{ll}
	6\times10^{-3}\\
~~~~~~~~{\rm for} \, n_{\rm gas} \lesssim 10^8 {\rm cm}^{-3} \\
2\times10^{-3} \left({n_\mathrm{gas} \over 10^{8} \mathrm{cm}^{-3}} \right)^{-1.0} 
	\left({\rho_\mathrm{BH} \over 10^6 M_\odot \mathrm{pc}^{-3}} \right)^{0.74} \\
~~~~~~~~{\rm for} \, n_{\rm gas} \gtrsim 10^8 {\rm cm}^{-3}. 
\end{array} 
\label{epsilon_c}
\right. 
\label{eq_bound}
\end{eqnarray}
The efficiency of $6\times 10^{-3}$ corresponds to a super-Eddington
accretion rate with the Eddington ratio of $\sim10$.  
The relation of equation (\ref{eq_bound}) 
can be basically understood by a condition from timescales
as $t_\mathrm{DF}\sim t_\mathrm{ac}$. 
Supposing $v\leq C\mathrm{s}$, 
this condition leads to $\epsilon_\mathrm{c}=\mathrm{const}$ in a low-density limit and
$\epsilon_\mathrm{c}\propto n_\mathrm{gas}^{-3/2}\rho_\mathrm{BH}$ in a high-density limit.
The difference in the power law of the critical condition between the numerical results and an analytic estimate
may come partially from mass accretion in supersonic phases and also from the gradual transition
from low-density regions to high-density regions.
Equation (\ref{eq_bound}) shows that if $\rho_\mathrm{BH}$ is lower ($r_\mathrm{typ}$ is larger), 
then the critical accretion efficiency becomes lower in high density environments of $n_{\rm gas} \gtrsim 10^8 {\rm cm}^{-3}$.
It is noted that as shown in \S\ref{mass-accretion-rate}, the accretion efficiency
is constrained by the radiation pressure if $\epsilon>0.1$.
However, the critical accretion efficiency given by (\ref{epsilon_c}) is considerably smaller
than $\epsilon=0.1$, and therefore is not affected
by the limitation of accretion rate due to the radiation pressure.

The present simulations show that
if the BH merger precedes the gas accretion, all BHs are likely to merge into
one large BH. This may be relevant to 
the fact that just one BH resides in the center of a massive galactic bulge.
On the other hand, if the gas accretion precedes the BH merger, 
multiple BHs are left in the system. 
Intriguingly, a lot of stellar mass ($\sim50 M_\odot$) BHs are found in the center of M31 \citep{bar14}. 
Although the origin of such a lot of stellar-mass BHs is not unveiled yet, 
multiple BHs can be left if the collective growth of a BH is driven by mass accretion.

\begin{figure}
	\centering
\includegraphics[width=85mm]{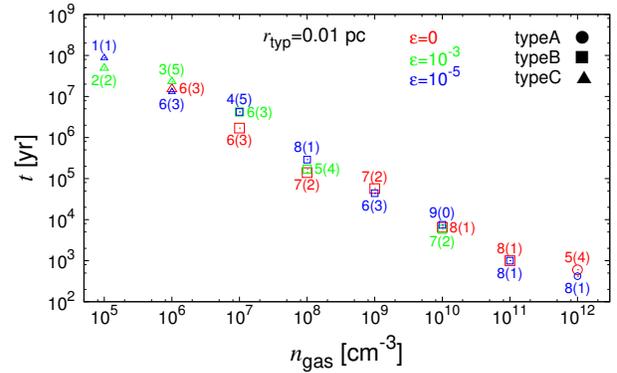}
	\caption{First merger time as a function of gas number density $n_\mathrm{gas}$ for $r_\mathrm{typ}=0.01~\mathrm{pc}$ 
		in the case that the merger dominantly contribute to the first growth of the BHs. 
		Red, green, and blue plots represent the merger time in the cases of no accretion ($\epsilon=0$), 
		intermediate accretion ($\epsilon=10^{-3}$), and lower accretion ($\epsilon=10^{-5}$), respectively. 
		Also, circle, square, and triangle symbols represent the types A, B, and C mergers, respectively. 
		The number (in parentheses) described beside each symbol represent the number of merged (escaped) BHs. 
	}
	\label{time_001}
\end{figure}

\subsection{Effects of Boundary Conditions}

In this paper, we have assumed the finite gas mass in a sphere. 
Here, we investigate the influences of this treatment on the merger timescale and mechanisms. 
In Fig. \ref{time_001}, we compare the merger timescale for no accretion ($\epsilon=0$, red symbols) to 
that for intermediate accretion ($\epsilon=10^{-3}$, green symbols) or for lower accretion ($\epsilon=10^{-5}$, blue symbols),  
in the cases in which the merger dominantly contributes to the first growth of BHs. 
In those cases, it is found that the boundary conditions of gas distributions do not bring 
significant influence on the merger timescale within uncertainties of a factor of $2$.
Furthermore, the merger mechanisms are also not influenced by the finiteness of the gas mass and the gas accretion. 

Since the total mass in the system is finite, escapers due to the slingshot mechanism emerge. 
The averaged number of escapers is $2\sim3$ in the present system with $M_\mathrm{gas, tot}=10^5~{M_\odot}$. 
The number of escaped BHs is expected to increase in a shallower gravitational potential. 
Therefore, in order for multiple BHs to merge via type B or C mechanism,
a deep potential is requisite.

\subsection{Effect of Recoil Kick}

\begin{figure}
	\hspace*{-0.5cm}
\includegraphics[width=90mm]{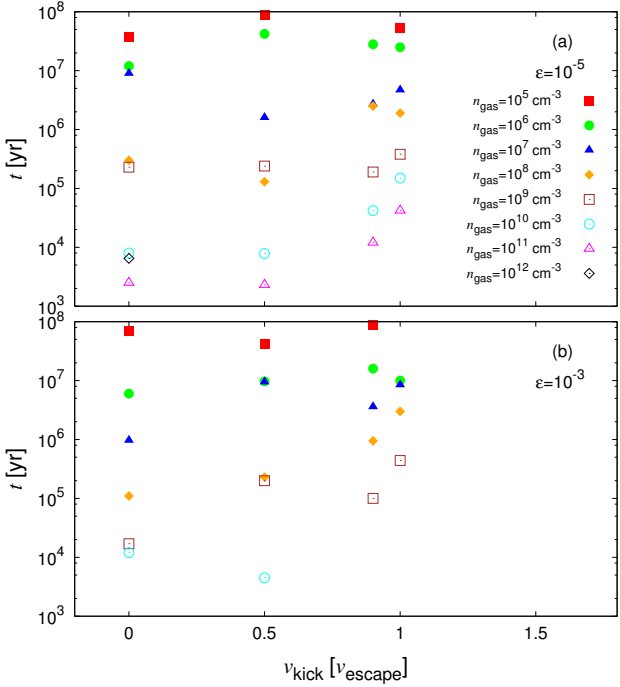}
	\caption{
		The averaged merger time as a function of the kick velocity in the cases of $r_\mathrm{typ}=0.01$ pc 
		and  $v_\mathrm{kick}\leq v_\mathrm{escape}$. 
		Filled red squares, green circles, blue triangles, orange diamonds, 
		open brown squares, blue circles, magenta triangles, and black diamonds 
		represent the merger time 
		for $n_\mathrm{gas}=10^5,~10^6,~10^7,~10^8,~10^9,~10^{10},~10^{11},$ and $10^{12}~
		\mathrm{cm}^{-3}
		$, respectively. 
		Panel (a) presents the results 
		for the case of lower accretion ($\epsilon =10^{-5}$), while  
		panel (b) for intermediate accretion ($\epsilon=10^{-3}$). 
	}
	\label{kick_35}
\end{figure}

In this section, we consider the effect of the recoil kick due to the anisotropic emission 
of gravitational wave radiation. 
The recoil velocity depends sensitively on the magnitudes of BH spins and 
the alignment with the orbital angular momentum of a BH binary.

To investigate the effect of the recoil kick, 
we perform the simulations incorporating the recoil in the case of $r_\mathrm{typ}=0.01$ pc.
In these simulations, 
we give the kick velocity $v_\mathrm{kick}$ as 0, 0.5, 0.9, 1.0, 1.1, and 1.5 times escape velocity
$v_\mathrm{escape}$.
Also, the number of BHs is set to be five to spare a computational cost. 
As expected, we find that if $v_\mathrm{kick}\leq1.0\times v_\mathrm{escape}$, 
the merged BH stays in the system, while  
in the cases of $v_\mathrm{kick}\geq1.1\times v_\mathrm{escape}$, 
the successive BH merger is halted due to the escape of the merged BH. 
In  Fig. \ref{kick_35},  we show the averaged merger time as a function of the kick velocity
for  $v_\mathrm{kick}\leq v_\mathrm{escape}$.
From this figure, the merger timescale for $v_\mathrm{kick}\sim v_\mathrm{escape}$
is almost the same as that for $v_\mathrm{kick}=0$ in low gas density cases  as $n_\mathrm{gas}< 10^{8}~\mathrm{cm}^{-3}$. 
This is because 
the merger timescale is almost irrespective of the initial extension of BH distributions in low BH density cases \citep{tag15}. 
On the other hand, in high gas density cases ($n_\mathrm{gas}\geq 10^{8}~\mathrm{cm}^{-3}$), 
the merger timescale for $v_\mathrm{kick} \sim v_\mathrm{escape}$ 
is longer by about one order of magnitude than that for $v_\mathrm{kick} = 0~\mathrm{km/s}$. 
From these results, the critical accretion rate is expected to be reduced by about one order of magnitude 
in high gas density regions, if the non-negligible recoil velocity is incorporated. 

\begin{figure}
	\hspace*{-0.5cm}
\includegraphics[width=90mm]{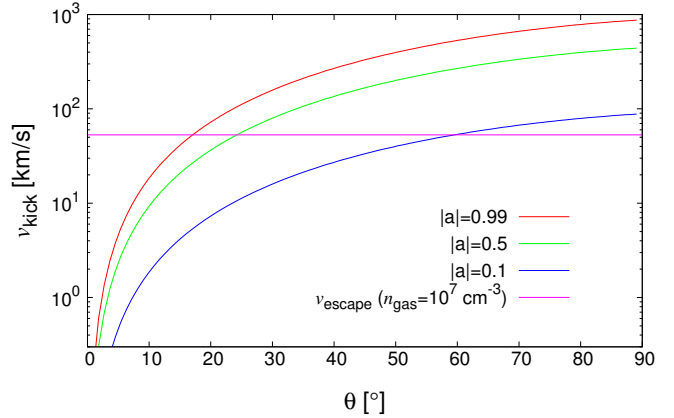}
	\caption{
		The averaged kick velocity as a function of the angle ($\theta$)
		between the spin of one BH and the angular momentum of the binary 
		in the cases that two BHs have the same masses and spin magnitude, 
		and the spin direction of another BH is the same as the direction of the angular momentum of the binary. 
		Red, green, and blue lines represent the results for BH spin magnitude of $|a_1|=|a_2|=$ 0.99, 0.5, and 0.1, respectively. 
		The pink line denotes the escape velocity for a gas cloud of $n_\mathrm{gas}=10^7~\mathrm{cm}^{-3}$. 
	}
	\label{kick_th}
\end{figure}

Next, we assess the probability that the kick velocity exceeds the escape velocity for each gas density. 
Here, we assume perfectly random orientation and magnitude of BH spins for
equal mass BHs. To derive the probability, we calculate $10^7$ sets of BH spins. 
The model of the recoil velocity is employed from \citet{cam07}, 
and we give the two angle $\xi$ and $\Theta_0$ in equation (1) of \citet{cam07} 
as $\xi=90^\circ$ and $\Theta_0=0^\circ$. 
In Table. \ref{kick_pro}, the escape velocity for each gas density 
and the probability of $v_\mathrm{kick}>v_\mathrm{escape}$ are shown.   
From this table, we expect that in a first-generation object ($n_\mathrm{gas}\sim10^7~\mathrm{cm}^{-3}$), 
about 90 percent of merged BHs may escape from the system. 
Thus, the effect of recoil velocity can reduce the BH merger rate significantly. 
This estimate is the results for the random directions and magnitudes of BH spins. 
On the other hand, 
if BH spins and the orbital angular momentum are aligned with each other, 
the probability may change. 
Considering the binary formation, 
we can suppose that the directions of BH spins are similar to the direction of the angular momentum of the binary. 
To make a simple estimate, we change the direction of the spin (${\bf a}_1$) of one BH, 
assuming that the direction of the other BH spin (${\bf a}_2$) 
is the same as the orbital angular momentum direction, 
and the spin magnitude is the same  ($|{\bf a}_1|=|{\bf a}_2|$). 
The resultant kick velocities are averaged over $10^5$ sets of 
the random azimuthal angles 
between 
the projected direction of ${\bf a}_1$ to the orbital plane of the binary and
the infall direction at a merger. 
Fig. \ref{kick_th} shows the kick velocity as a function of $\theta$. 
As a result, in the cases of the rapidly rotating BHs ($|{\bf a}_1|=|{\bf a}_2|=0.99$) mergers, 
the averaged kick velocity exceeds the escape velocity in $\theta >16.2^\circ$. 
This means that if $\theta$ is randomly given from $0^\circ$ to $90^\circ$, 
about 82 percent will escape and 18 percent stay in the system.  
According as the BH spin magnitude decreases, the escape fraction decreases. 
Thus, it is expected that the escape probability of a merged BH stemming form 
a binary becomes lower than that for the cases of the random spins. 
Furthermore, \citet{nat98} have suggested 
that the spins of BHs are aligned with the outer accretion disk typically within $\sim 10^5-10^6$ yr 
due to the Bardeen-Petterson effect, although
the large uncertainty in the viscosity may vary the alignment timescale \citep{vol07}. 

\begin{table}
	\vspace{0mm}
	\caption{The escape velocity and probability for each gas number density in the cases of random spins and equal mass BHs.}
\label{kick_pro}
\begin{tabular}{c|c|c}
\hline
$n_\mathrm{gas}$ ($\mathrm{cm}^{-3}$)&
$v_\mathrm{escape}$ (km/s)&
$P(v_\mathrm{kick}>v_\mathrm{escape})~(\%)$
\\
\hline
$10^5$&24.6&95.6\\
\hline
$10^6$&36.1&93.4\\
\hline
$10^7$&53.0&90.2\\
\hline
$10^8$&77.7&85.3\\
\hline
$10^9$&114&78.0\\
\hline
$10^{10}$&167&67.2\\
\hline
$10^{11}$&246&52.5\\
\hline
$10^{12}$&361&35.8\\
\hline
\end{tabular}
\end{table}

\subsection{Merger in GW150914}

\begin{figure}
\begin{center}
\includegraphics[width=85mm]{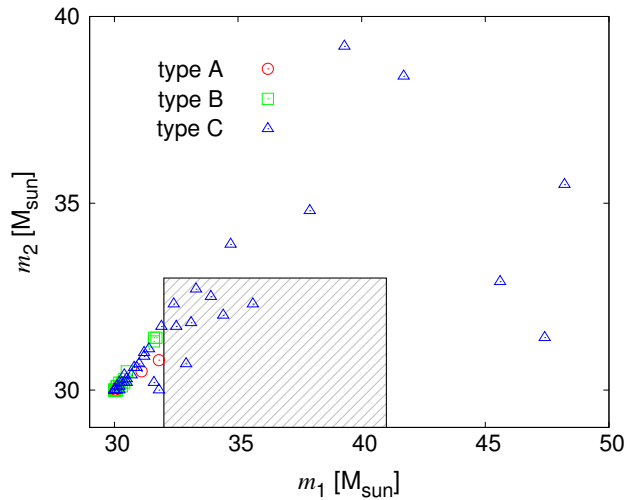}
\vspace{-5mm}
 \caption{The masses in a binary just before the first merger in each run. 
	 Red circle, green square and blue triangle plots represent type A, B and C mergers. 
	 The masses inferred from GW150914 with their uncertainties are indicated by the hatched regions. 
 }
\label{mass_multi_1st}
\end{center}
\end{figure}

Very recently, the Laser Interferometer Gravitational-Wave Observatory (LIGO)
has detected the gravitational wave event,  GW150914, as a result of the merger of a $\sim 30~M_\odot$ BH binary  
\citep{abbott16}. 
This binary is composed of a $36^{+5}_{-4}M_\odot$ and  $29^{+4}_{-4}M_\odot$ BH, and
the final BH mass is $62^{+4}_{-4}M_\odot$. 
These observations demonstrate that binary stellar-mass black hole systems exist
and they can merge within the cosmic time. 
The masses of a binary BH in the GW150914 event are close to our assumption of 
the initial BH mass. 
\footnote{After the present calculations were almost completed, 
the detection of gravitational waves in GW150914 was reported. 
Hence, we have not explored whether BHs with initial mass lower than $30~M_\odot$ 
can account for the BH binary in GW150914. This will be studied in a forth-coming paper.}
Based on the present numerical simulations, obviously the merger event 
in GW150914 is not an accretion-driven merger,
since the final mass is not so high as expected in an accretion-driven merger. 
In Figure \ref{mass_multi_1st}, the masses in a binary just before the first merger
in our simulations are shown. We find that in several sets of parameters,
the masses in a binary match those in GW150914 within their uncertainties.
All the mergers in these parameters are driven by three-body interactions (type C) 
in the range of gas density from $n_{\rm gas}=10^5~{\rm cm}^{-3}$ to $10^{10}~{\rm cm}^{-3}$.
The density of $n_{\rm gas}=10^{5-8}~{\rm cm}^{-3}$ corresponds to
high density interstellar clouds, and $n_{\rm gas} \gtrsim 10^{8}~{\rm cm}^{-3}$ corresponds
to the galactic center regions of $\lesssim 1~{\rm pc}$. 
Also, these three body-driven mergers are accompanied by highly super-Eddington accretion,
and the mass of a few $M_\odot$ is accreted before mergers. 
Such gas accretion might produce electromagnetic emission associated with
gravitational wave events.  
\footnote{It was reported that 
a weak transient source above 50 keV was detected 0.4 s after the GW150914 event 
by Fermi Gamma-ray Burst Monitor (GBM) \citep{Connaughton16}.  However, it is pointed out
that the GBM transient event is very unlikely associated with the GW150914
\citep{Greiner16,Xiong16}.}

\section{CONCLUSIONS}

In this paper, we have explored the merger of multiple BHs systems under background gas
distributions, incorporating dynamical friction and gas accretion. 
For the purpose, we have performed highly accurate post-Newtonian
numerical simulations taking into account such general relativistic effects as
the pericenter shift and gravitational wave emission. 
Consequently, we have found the followings. 

\begin{enumerate}
		\renewcommand{\labelenumi}{(\arabic{enumi})}

\item \noindent
The dominant mechanisms for the growth of BHs are varied 
according to the gas accretion rate $\dot{m}$, the BH density $\rho_\mathrm{BH}$, 
and the gas number density $n_\mathrm{gas}$. 
The merger mechanisms are classified into a gas drag-driven merger (type A), 
an interplay-driven merger (type B), a three body-driven merger (type C), 
or an accretion-driven merger (type D). 
\\

\item \noindent
In a type A, B, or C merger, all BHs can merge eventually into one heavy BH, 
if the BH merger precedes the gas accretion. 
However, in a type D, the merger of all BHs into one BH is hard to occur,
leaving several BHs around a primary BH. 
\\

\item \noindent
We have derived the critical accretion rate $\dot{m}_{\rm c}$, 
below which the BH growth is predominantly promoted by mergers. 
The $\dot{m}_{\rm c}$ is fitted as 
\begin{eqnarray}
	\frac{\dot{m}_{\rm c}}{\dot{m}_{\rm HL}}=
\left \{
\begin{array}{ll}
	6\times10^{-3}\\
~~~~~~~~{\rm for} \, n_{\rm gas} \lesssim 10^8 ~{\rm cm}^{-3} \\
2\times10^{-3} \left({n_\mathrm{gas} \over 10^{8} ~\mathrm{cm}^{-3}} \right)^{-1.0} 
	\left({\rho_\mathrm{BH} \over 10^6 ~M_\odot \mathrm{pc}^{-3}} \right)^{0.74} \\
~~~~~~~~{\rm for} \, n_{\rm gas} \gtrsim 10^8 ~{\rm cm}^{-3}. 
\end{array}
\right. 
\end{eqnarray}
Note that the effect of the recoil kick may reduce the critical accretion rate 
by about one order of magnitude for high gas density regions. 
Also, since our analytic prescriptions of the gas effects do not take the inhomogeneity of gas distributions
into account, our simulations may overestimate the effects of the mass accretion and the dynamical friction by gas. 
\\

\item \noindent
The recoil kick due to the anisotropic emission of gravitational wave radiation is 
important to estimate the event rate of BH mergers. 
We have estimated that roughly ninety percent of merged BHs can escape
from a first-generation object, 
if 
the directions and magnitudes of BH spins are 
completely random. The escape probability is reduced, if BH spins and
the orbital angular momentum are aligned with each other.
\\

\item \noindent
Supposing the gas and BH density based on the recent simulations on the first star formation, 
the BH merger proceeds before the significant mass accretion,  
if the accretion rate is lower than $\sim10$ Eddington accretion rate. 
\\

\item \noindent
The merger of a binary BH system in GW150914 is most
likely to be driven by three-body encounters accompanied by a few $M_\odot$ of gas accretion. 

\end{enumerate}

In this paper, we do not consider the dynamics of the gas. 
However, the effect of the gas dynamics may affect the merger or accretion mechanism. 
We will investigate the influence of the effect of the gas dynamics on the BH merger in the future. 

\section*{Acknowledgments}

We thank Taihei Yano for valuable discussions. 
Numerical computations and analyses were carried out on Cray XC30 and computers at Center for Computational Astrophysics, 
National Astronomical Observatory of Japan, respectively. 
This research was also supported in part by Interdisciplinary Computational Science Program in Center for Computational Sciences, 
University of Tsukuba,
and Grant-in-Aid for Scientific Research (B) by JSPS (15H03638).








\begin{thebibliography}{99}
	\bibitem[\protect\citeauthoryear{Aasi et al.}{2013}]{aasi13} Aasi J., et al., 2013, Phys. Rev. D., 87, 022002 
	\bibitem[\protect\citeauthoryear{Abbott et al.}{2006}]{abbott06} Abbott B. P., et al., 2006, Phys. Rev. D., 73, 062001 
	\bibitem[\protect\citeauthoryear{Abbott et al.}{2016}]{abbott16} Abbott B. P., et al., 2016, Phys. Rev. Lett., 116, 061102
	\bibitem[\protect\citeauthoryear{Abbott et al.}{2016}]{abbott16b} Abbott B. P., et al., 2016, ApJ, 818, L22 
	\bibitem[\protect\citeauthoryear{Abramowicz et al.}{1988}]{Abramowicz88} Abramowicz M.~A., Czerny B., Lasota J.~P., Szuszkiewicz E., 1988, ApJ, 332, 646 
	\bibitem[\protect\citeauthoryear{Alvarez, Wise, \& Abel}{2009}]{alv09} Alvarez M. A., Wise J.~H., Abel T., 2009, ApJ, 701, L133 
	\bibitem[\protect\citeauthoryear{Aso et al.}{2013}]{Aso13} Aso Y., Michimura Y., Somiya K., Ando M., Miyakawa O., Sekiguchi T., 
Tatsumi D., Yamamoto H., 2013, PhRvD, 88, 043007 
	\bibitem[\protect\citeauthoryear{Barnard et al.}{2014}]{bar14} Barnard R., Garcia M. R., Primini F. Murray S. S., 2014, ApJ, 791, 33
		\bibitem[\protect\citeauthoryear{Bergin, Snell, \& Goldsmith}{1996}]{Bergin96} Bergin E.~A., Snell R.~L., Goldsmith P.~F., 1996, ApJ, 460, 343 
	\bibitem[\protect\citeauthoryear{Berti, Cardoso, \& Will}{2006}]{Berti06} Berti E., Cardoso V., Will C.~M., 2006, Phys. Rev. D., 73, 064030 
	\bibitem[\protect\citeauthoryear{Bondi \& Hoyle}{1944}]{bon44} Bondi H., Hoyle F., 1944, MNRAS, 104, 273
	\bibitem[\protect\citeauthoryear{Campanelli et al.}{2007}]{cam07} Campanelli M., Lousto C. O., Zlochower Y., Merritt D., 2007, PhRvL, 98, 1102
	\bibitem[\protect\citeauthoryear{Connaughton et al.}{2016}]{Connaughton16} Connaughton V., et al., 2016, arXiv, arXiv:1602.03920 
	\bibitem[Djorgovski et al.(2007)]{Djorgovski07} Djorgovski~S.~G., Courbin~F., Meylan~G., Sluse~D., Thompson~D., Mahabal~A., Glikman~E., 2007, ApJ, 662, L1
	\bibitem[\protect\citeauthoryear{Escala et al.}{2004}]{esc04} Escala A., Larson R., Coppi P., Mardones D., 2004, ApJ, 507, 765
	\bibitem[\protect\citeauthoryear{Escala et al.}{2005}]{esc05} Escala A., Larson R., Coppi P., Mardones D., 2005, ApJ, 630, 152
	\bibitem[\protect\citeauthoryear{Fan et al.}{2001}]{fan01} Fan X. et al., 2001, AJ, 122, 2833
	\bibitem[Farina et al.(2013)]{Farina13} Farina~E.~P., Montuori~C., Decarli~R., Fumagalli~M., 2013, MNRAS, 431, 1019
	\bibitem[\protect\citeauthoryear{Greif et al.}{2011}]{gre11} Greif T.~H., Springel V., White S.~D.~M., Glover S.~C.~O., Clark P.~C., Smith R.~J., Klessen R.~S., Bromm V., 2011, ApJ, 737, 75 
\bibitem[\protect\citeauthoryear{Greiner et al.}{2016}]{Greiner16} Greiner J., Burgess J.~M., Savchenko V., Yu H.-F., 2016, arXiv, arXiv:1606.00314 
	\bibitem[\protect\citeauthoryear{Haiman}{2013}]{hai13} Haiman Z., 2013, in Wiklind T., Mobasher B., Bromm V., eds, Astrophysics and Space Science Library, Vol. 396, The First Galaxies. Springer-Verlag, Berlin, p. 293
	\bibitem[\protect\citeauthoryear{Hanamoto, Ioroi \& Fukue}{2001}]{han01} Hanamoto K., Ioroi M., Fukue J., 2001, PASJ, 53, 105
	\bibitem[\protect\citeauthoryear{Heger \& Woosley}{2002}]{heg02} Heger A., Woosley S. E., 2002, ApJ, 567, 532
	\bibitem[\protect\citeauthoryear{Hirano et al.}{2014}]{Hirano14} Hirano S., Hosokawa T., Yoshida N., Umeda H., Omukai K., Chiaki G., Yorke H.~W., 2014, ApJ, 781, 60 
	\bibitem[\protect\citeauthoryear{Hoyle \& Lyttleton}{1939}]{hoy39} Hoyle F., Lyttleton R. A., 1939, Proc. Camb. Phil. Soc., 35, 405
	\bibitem[\protect\citeauthoryear{Iwasawa, Funato \& Makino}{2006}]{iwa06}Iwasawa M., Funato Y., Makino J., 2006, ApJ, 651, 1059,
	\bibitem[\protect\citeauthoryear{Kormendy \& Ho}{2013}]{kor13} Kormendy J., Ho L.~C., 2013, ARA\&A, 51, 511 
	\bibitem[\protect\citeauthoryear{Kupi}{2006}]{kup06} Kupi G., Amaro-Seoane P., Spurzem R., 2006, MNRAS, 371,45
	\bibitem[\protect\citeauthoryear{Kurk et al.}{2007}]{kur07} Kurk J.~D., et al., 2007, ApJ, 669, 32 
	\bibitem[Liu, Shen \& Strauss(2011)]{Liu11} Liu~X., Shen~Y.,Strauss~M.~A., 2011, ApJ, 736, L7
	\bibitem[\protect\citeauthoryear{L{\"u}ck et al.}{2006}]{Huck06} L{\"u}ck H., et al., 2006, CQGra, 23, S71 
	\bibitem[\protect\citeauthoryear{Makino \& Aarseth}{1992}]{mak92} Makino J. Aarseth S., 1992, PASJ, 44, 141
	\bibitem[\protect\citeauthoryear{Merritt \& Poon}{2004}]{mer04} Merritt~D., Poon~M.~Y., 2004, ApJ, 606, 788
	\bibitem[\protect\citeauthoryear{Milosavljevic, Couch \& Bromm}{2009}]{mil09} Milosavljevic M., Couch S. M., Bromm V., 2009, ApJ, 696, 146
	\bibitem[\protect\citeauthoryear{Mortlock et al.}{2011}]{mor11} Mortlock D.~J., et al., 2011, Nature, 474, 616 
	\bibitem[\protect\citeauthoryear{Namekata \& Umemura}{2016}]{NU16} Namekata D., Umemura M., 2016, MNRAS, 460, 980 
	\bibitem[\protect\citeauthoryear{Natarajan \& Pringel et al.}{1998}]{nat98} Natarajan P., Pringele J. E., 1998, ApJ, 506, L97
	\bibitem[\protect\citeauthoryear{Omukai}{2000}]{omu00} Omukai K., 2000, ApJ, 534, 809
	\bibitem[\protect\citeauthoryear{Ostriker}{1999}]{ost99} Ostriker E. C., 1999, ApJ, 513, 252
	\bibitem[Schawinski et al.(2011)]{Schawinski11} Schawinski~K.,Urry~M., Treister~E., Simmons~B., Natarajan~P., Glikman~E., 2011, ApJ, 743, L37
	\bibitem[\protect\citeauthoryear{Sesana et al.}{2005}]{Sesana05} Sesana A., Haardt F., Madau P., Volonteri M., 2005, ApJ, 623, 23 
	\bibitem[\protect\citeauthoryear{Susa}{2013}]{sus13} Susa H., 2013, ApJ, 773, 185
	\bibitem[\protect\citeauthoryear{Susa, Hasegawa, \& Tominaga}{2014}]{sus14} Susa H., Hasegawa K., Tominaga N., 2014, ApJ, 792, 32 
	\bibitem[\protect\citeauthoryear{Tagawa et al.}{2015}]{tag15} Tagawa H., Umemura M., Gouda N., Yano T., Yamai Y., 2015, MNRAS, 451, 2174
	\bibitem[\protect\citeauthoryear{Tanaka \& Haiman}{2009}]{tan09} Tanaka T., Haiman Z., 2009, A \& A, 696, 1798
	\bibitem[\protect\citeauthoryear{Tanikawa \& Umemura}{2011}]{tan11} Tanikawa A., Umemura M., 2011, MNRAS, 728, 31
	\bibitem[\protect\citeauthoryear{Tanikawa \& Umemura}{2014}]{tan14} Tanikawa A., Umemura M., 2014, MNRAS, 440, 652
	\bibitem[\protect\citeauthoryear{Umemura}{2001}]{ume01} Umemura M., 2001, ApJ, 560, L29
	\bibitem[\protect\citeauthoryear{Umemura, Loeb, \& Turner}{1993}]{ume93} Umemura M., Loeb A., Turner E.~L., 1993, ApJ, 419, 459 
	\bibitem[\protect\citeauthoryear{Umemura et al.}{2012}]{ume12} Umemura M., Susa H., Hasegawa K., Suwa T., Semelin B., 2012, PTEP, 2012, 01A306 
	\bibitem[\protect\citeauthoryear{Volonteri \& Bellovary}{2012}]{vol12} Volonteri M., Bellovary J., 2012, Rep. Prog. Phys., 75, 124901 
	\bibitem[\protect\citeauthoryear{Volonteri \& Rees}{2005}]{vol05} Volonteri M., Rees M. J., 2005, ApJ, 633, 624
	\bibitem[\protect\citeauthoryear{Volonteri et al.}{2007}]{vol07} Volonteri M., Sikora M., Lasota J.-P., 2007, ApJ, 667, 704
	\bibitem[\protect\citeauthoryear{Volonteri, Silk \& Dubus}{2015}]{vol15} Volonteri M., Silk J., Dubus G., 2015, ApJ, 804, 148
	\bibitem[\protect\citeauthoryear{Watarai et al.}{2000}]{wat00} Watarai K., Fukue J., Takeuchi M., Mineshige S., 2000, PASJ, 52, 133
	\bibitem[\protect\citeauthoryear{Wu et al.}{2015}]{wu15} Wu X.-B., et al., 2015, Nature, 518, 512 
	\bibitem[\protect\citeauthoryear{Wyithe \& Loeb}{2003}]{WL03} Wyithe J.~S.~B., Loeb A., 2003, ApJ, 590, 691
	\bibitem[\protect\citeauthoryear{Xiong}{2016}]{Xiong16} Xiong S., 2016, arXiv, arXiv:1605.05447 
	\bibitem[\protect\citeauthoryear{Yajima et al.}{2015}]{Yajima15} Yajima H., Li Y., Zhu Q., Abel T., 2015, ApJ, 801, 52 
\end{thebibliography}
\end{document}